\documentclass[aps,prl,reprint,amsmath,amssymb,superscriptaddress]{revtex4-2}
\usepackage{graphicx}
\graphicspath{{../pic/}}
\usepackage{epstopdf}
\usepackage{dcolumn}
\usepackage{bm}
\usepackage{booktabs}
\begin{document}
\title{Emergent Electric-Field-Tunable $p$-Wave Magnetism in Twisted Bilayer NiPS$_3$}
\author{Jiaran Su}
\affiliation{State Key Laboratory of High Pressure and Superhard Materials, Key Laboratory of Material Simulation Methods \& Software of Ministry of Education, College of Physics, Jilin University}
\affiliation{Tsientang Institute for Advanced Study, Zhejiang 310024, China}

\author{Jin-Xin Hu}
\email{jxhu@tias.ac.cn}
\affiliation{Tsientang Institute for Advanced Study, Zhejiang 310024, China}
\author{Qi Sun}
\affiliation{State Key Laboratory of High Pressure and Superhard Materials, Key Laboratory of Material Simulation Methods \& Software of Ministry of Education, College of Physics, Jilin University}
\affiliation{Tsientang Institute for Advanced Study, Zhejiang 310024, China}
\author{Dante M. Kennes}
\affiliation{Institut f\"ur Theorie der Statistischen Physik, RWTH Aachen University, Aachen, Germany}
\affiliation{JARA -- Fundamentals of Future Information Technology, Aachen, Germany}
\affiliation{Max Planck Institute for the Structure and Dynamics of Matter, Center for Free-Electron Laser Science, Hamburg, Germany}
\author{Yuhao Fu}
\email{fuyuhaoy@gmail.com}
\affiliation{State Key Laboratory of High Pressure and Superhard Materials, Key Laboratory of Material Simulation Methods \& Software of Ministry of Education, College of Physics, Jilin University}
\author{Lede Xian}
\email{ldxian@tias.ac.cn}
\affiliation{Tsientang Institute for Advanced Study, Zhejiang 310024, China}
\affiliation{Max Planck Institute for the Structure and Dynamics of Matter, Center for Free-Electron Laser Science, Hamburg, Germany}
\affiliation{Songshan Lake Materials Laboratory, 523808 Dongguan, Guangdong, China}
\author{Lijun Zhang}
\email{lijun\_zhang@jlu.edu.cn}
\affiliation{State Key Laboratory of Integrated Optoelectronics, Key Laboratory of Automobile Materials of MOE,
College of Materials Science and Engineering, Jilin University, Changchun 130012, China}

\date{\today}
\begin{abstract}
A perpendicular electric field can reorient a momentum-space spin texture without changing the local magnetic configuration. In this Letter, we investigate this mechanism in a $21^\circ$-twisted $\mathrm{NiPS}_3$ bilayer with zigzag order and coplanar, nonparallel layer N\'eel vectors. In the ideal nonrelativistic limit, an antiunitary in-plane half-translation enforces opposite-spin pairing at opposite momenta, while a layer-exchanging twofold operation pins the leading $p$-wave spin splitting to a crystallographic direction. A perpendicular layer bias lifts this orientational constraint while preserving the antiunitary symmetry, allowing a transverse component of the splitting vector. First-principles calculations show gate-dependent reorientation and deformation of spin-resolved isoenergy contours. Wannier-based linear-response calculations further reveal field-activated diagonal spin conductivities and a polarity-dependent tilt in the angular profile of the transverse spin-current response. The spin-texture axis, dispersion axes, and transport-response axis are distinct quantities governed by the same symmetry reduction. These results identify layer bias as a route to controlling odd-parity spin textures and spin-current anisotropy, provided that the coplanar magnetic state remains stable.
\end{abstract}
\maketitle

\emph{Introduction.---}Electrical generation of spin currents is central to spintronics, but device applications also require control over the direction and polarization of these currents~\cite{sinova2015,manchon2019}. Crystal and magnetic symmetries constrain this directional response, as illustrated by unconventional spin--orbit torques in low-symmetry $\mathrm{WTe}_2$~\cite{macneill2017}. Exchange-driven spin splitting in compensated magnets offers a complementary route to spin-dependent electronic motion even in the absence of strong spin--orbit coupling~\cite{gonzalez2021,gonzalez2024}. An important question is whether an electrostatic gate can continuously reorient the momentum-space anisotropy and the resulting spin-current response while the local magnetic configuration remains fixed.

Compensated magnets provide several symmetry-distinct settings for this problem. Collinear altermagnets exhibit even-parity, momentum-dependent exchange splitting~\cite{smejkal2022prx,smejkal2022landscape,ma2021svl,yuan2020,hayami2020bottom,smejkal2020hall,gonzalez2021,smejkal2022gmr,krempasky2024,reimers2024direct,amin2024mnte,zhou2025crsb,song2025nrm,liu2024twist,pan2024stack}, whereas coplanar $p$-wave magnets can exhibit odd-parity spin polarization associated with time reversal combined with a translation~\cite{brekke2024,hellenes2023,yu2025,songz2025,yamada2025metallic,ma2026hyperspin,sim2026quantum,leeb2026collinear}. Odd parity relates spin polarizations at opposite momenta and is associated with phenomena such as a nonrelativistic Edelstein response~\cite{ezawa2025out,ezawa2025,chakraborty2025nree} and superconducting proximity effects~\cite{fukaya2025superconducting,khodas2026nonrelativistic,sukhachov2025coexistence,sun2025pseudo}. Additional spatial symmetries determine whether the texture axis is pinned to the lattice. These distinct symmetry requirements suggest a route to electrically reorienting the texture while preserving its odd parity.

Twisted van der Waals magnets offer a suitable platform because interlayer rotation changes crystalline and magnetic symmetries, while a perpendicular electric field controls layer imbalance~\cite{zhang2024slc,song2021moire,liu2024twist}. Twisting has already been proposed to generate nonrelativistic spin splitting and transverse spin transport in antiferromagnetic bilayers~\cite{he2023,liu2024twist,sheoran2024}. Electrical control also extends to odd-parity magnets. Coupling between ferroelectric polarization and magnetic chirality enables switching in $\mathrm{NiI}_2$~\cite{song2025pwave}, and antiferromagnet/metal/antiferromagnet heterostructures have been proposed as a means of controlling a $p$-wave phase and its Edelstein response through carrier filling~\cite{kim2026pwave}. These developments raise the question of whether a layer potential can control an existing odd-parity texture without changing the underlying magnetic configuration.

Motivated by the predominantly in-plane moments of $\mathrm{NiPS}_3$~\cite{wildes2015}, we investigate this mechanism in a $21^\circ$-twisted bilayer with zigzag order and nonparallel, coplanar layer N\'eel vectors. Our results show that, although the monolayer band structure is spin degenerate, the twisted system exhibits a $p$-wave spin-splitting texture. In the nonrelativistic symmetry description, an in-plane half-translation combined with time reversal preserves opposite-spin pairing at opposite momenta, whereas a layer-exchanging twofold operation pins the texture axis at zero gate field. A perpendicular electric field breaks the twofold symmetry while preserving the antiunitary half-translation. The leading $p$-wave splitting can consequently acquire a transverse momentum component, rotating its axis while retaining odd parity. First-principles calculations reveal the corresponding reorientation and deformation of spin-resolved contours, while Wannier-based linear-response calculations predict a polarity-dependent tilt in the angular spin-current response. These results identify layer bias as a means of releasing the orientational constraint over the field range in which the coplanar magnetic state remains stable.

\begin{figure}
    \centering
    \includegraphics[width=\columnwidth]{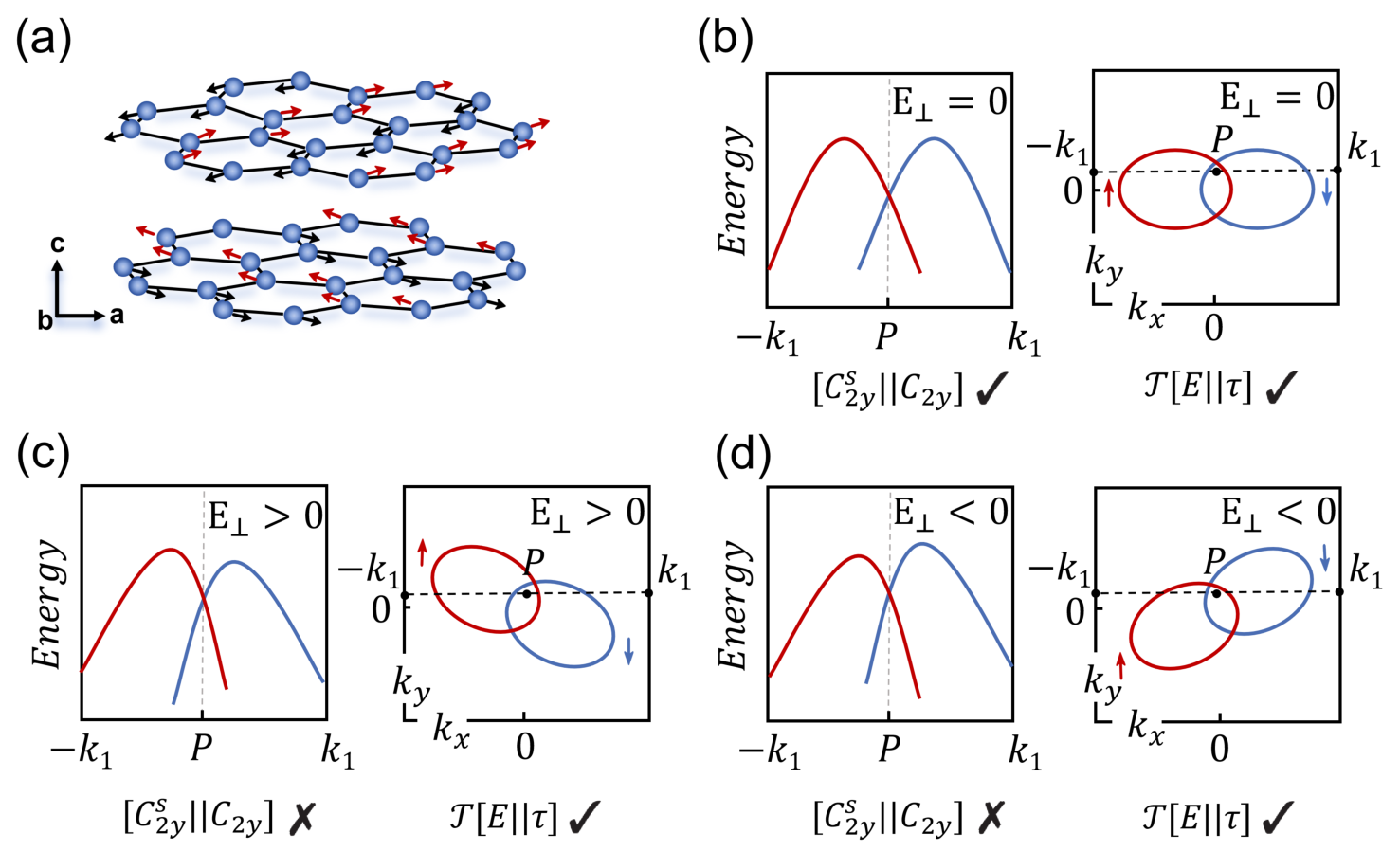}
     \caption{Schematic electric-field-induced reorientation of the spin texture in a twisted antiferromagnetic bilayer with coplanar N\'{e}el order. (a) Nonparallel in-plane moments define a finite vector chirality and remain fixed for all fields considered.
(b) At $E_\perp=0$, both $\mathcal{A}=\mathcal{T}\{E|\boldsymbol{\tau}\}$ and $\mathcal{B}=[C_{2y}^{s}\Vert C_{2y}]$ are preserved, and opposite-spin contours are displaced along $\pm k_x$.
The left panel shows the offset cut $k_y=k_y^0\neq0$, indicated by the horizontal dashed line in the right panel; $P=(0,k_y^0)\neq\Gamma$.
(c), (d) Positive and negative fields break $\mathcal{B}$, allowing opposite contour tilts and band asymmetries along this cut.
The surviving $\mathcal{A}$ pairs the $\pm k_y^0$ cuts and protects degeneracy at $\Gamma$, outside the displayed cut.
Red and blue denote opposite spin polarizations. The sketches illustrate symmetry relations rather than a quantitative correspondence between the displayed dispersions and contours.}
    \label{fig:schematic}
    \end{figure}

Figure~\ref{fig:schematic}(a) shows a coplanar noncollinear spin configuration formed by the N\'eel vectors of the two layers. Twisting changes the relative orientations of the two lattices and their magnetic anisotropy axes. The anisotropy and interlayer exchange jointly determine the angle between the moments~\cite{song2021moire,he2023,gonzalez2025mnps3}. The vector chirality $\boldsymbol{\chi}=\mathbf{n}_1\times\mathbf{n}_2$ characterizes this noncollinear order and vanishes in the collinear limit. In the two-order mechanism considered here, the nonrelativistic spin splitting normal to the spin plane vanishes in the collinear limit~\cite{hellenes2023,yu2025,hayami2020antisym}. For finite chirality, the full spin space group constrains both the spin polarization of the bands and its momentum dependence~\cite{litvin1977,smejkal2022prx,xiao2024ssg,chen2024ssg,jiang2024ssg}. Our analysis starts from the pair of $p$-wave branches shown in the figure. We neglect spin--orbit coupling and consider fields that preserve the coplanar configuration.

\emph{Symmetry analysis.---}The coplanar magnetic order preserves the antiunitary half-translation $\mathcal{A}=\mathcal{T}\{E|\boldsymbol{\tau}\}$, where $\mathcal{T}$ reverses all three spin components, $E$ denotes the identity spatial rotation, and $\boldsymbol{\tau}=(\tfrac12,\tfrac12,0)$ is an in-plane half-translation that exchanges the two magnetic sublattices. The net magnetization therefore vanishes. Time reversal maps $\mathbf{k}$ to $-\mathbf{k}$, and the translation contributes a Bloch phase $e^{-i\mathbf{k}\cdot\boldsymbol{\tau}}$. Thus $\mathcal{A}$ maps $(\mathbf{k},\uparrow)$ to a state $(-\mathbf{k},\downarrow)$ of the same energy. Here $\uparrow$ and $\downarrow$ denote opposite signs of the spin expectation value along $z$, the direction normal to the spin plane~\cite{hellenes2023,xiao2024ssg,jiang2024ssg}. Combining $\mathcal{A}$ with the coplanar spin-only antiunitary operation $[C_{2z}^{s}\mathcal{T}\Vert E]$ gives the unitary operation $\{C_{2z}^{s}\Vert E|\boldsymbol{\tau}\}$. This operation reverses $S_x$ and $S_y$ at fixed $\mathbf{k}$, so their expectation values vanish in a nondegenerate Bloch state. The split bands can therefore carry spin polarization only along the $z$ direction, consistent with the symmetry analysis of coplanar noncollinear magnets in Ref.~\cite{ma2026hyperspin}.

At zero field, the operation $\mathcal{B}=[C_{2y}^{s}\Vert C_{2y}]$ also exchanges the two layers. Its spatial part maps $(x,y,z)$ to $(-x,y,-z)$ and $(k_x,k_y)$ to $(-k_x,k_y)$, while its spin part reverses $S_z$. It therefore relates states with opposite spin polarizations at symmetry-related momenta~\cite{hellenes2023}. Such a relation can give rise to spin--valley locking in collinear antiferromagnets~\cite{ma2021svl}. In the present system, it pins the leading odd-parity spin splitting to the crystallographic $k_x$ direction. The operations labeled $\mathcal{T}\{E|\boldsymbol{\tau}\}$ and $[C_{2y}^{s}\Vert C_{2y}]$ in Fig.~\ref{fig:schematic} correspond to $\mathcal{A}$ and $\mathcal{B}$, respectively.

We label the two branches by their opposite spin polarizations and track each branch continuously through the crossing. We write $\varepsilon_\pm=\varepsilon_0\pm\Delta_p$, where $2\Delta_p$ is the signed spin splitting. At zero field, $\mathcal{A}$ and $\mathcal{B}$ operations impose
\begin{align}
\Delta_p(-\mathbf{k},0)&=-\Delta_p(\mathbf{k},0),\\
\Delta_p(-k_x,k_y,0)&=-\Delta_p(k_x,k_y,0).
\label{eq:zero}
\end{align}
These relations require the signed splitting to be odd in $k_x$ and even in $k_y$. Thus the leading $p$-wave term is proportional to $k_x$, and the branches are degenerate along $k_x=0$, including at $\Gamma$ [Fig.~\ref{fig:schematic}(b)]. The odd parity follows from $\mathcal{A}$, while $\mathcal{B}$ fixes the zero-field orientation.

A perpendicular electric field breaks the layer-exchanging symmetry $\mathcal{B}$ but preserves $\mathcal{A}$, whose spatial part is an in-plane translation~\cite{sivadas2016,zhang2024slc,qi2024slc,peng2025layer}. The relation $\Delta_p(-\mathbf{k},E_\perp)=-\Delta_p(\mathbf{k},E_\perp)$ therefore holds at finite field and protects the crossing at $\Gamma$. The operation $\mathcal{B}$ still relates the systems at opposite field polarities through
\begin{equation}
\Delta_p(k_x,k_y,E_\perp)
=-\Delta_p(-k_x,k_y,-E_\perp).
\label{eq:map}
\end{equation}
A transverse component of the $p$-wave splitting is therefore allowed and reverses sign with the field. When this component is finite, the nodal line tilts away from $k_x=0$ in opposite directions at opposite polarities [Figs.~\ref{fig:schematic}(c) and \ref{fig:schematic}(d)]. Combining $\mathcal{A}$ and $\mathcal{B}$ further yields
\begin{equation}
\varepsilon_s(k_x,k_y;E_\perp)
=\varepsilon_s(k_x,-k_y;-E_\perp),
\label{eq:mirror}
\end{equation}
so the isoenergy contours of a given spin branch at opposite field polarities are related by reflection across the $k_x$ axis when compared at the same energy. The field can also modify the band curvature, changing both the orientation and the shape of the contours. The spin-splitting axis and the principal axes of the dispersion can therefore respond differently to the field while the underlying N\'eel vectors and their chirality remain fixed~\cite{sivadas2016,huang2018,jiang2018nmat,ma2021svl}.

\begin{figure}[tbp]
    \centering
    \includegraphics[width=\columnwidth]{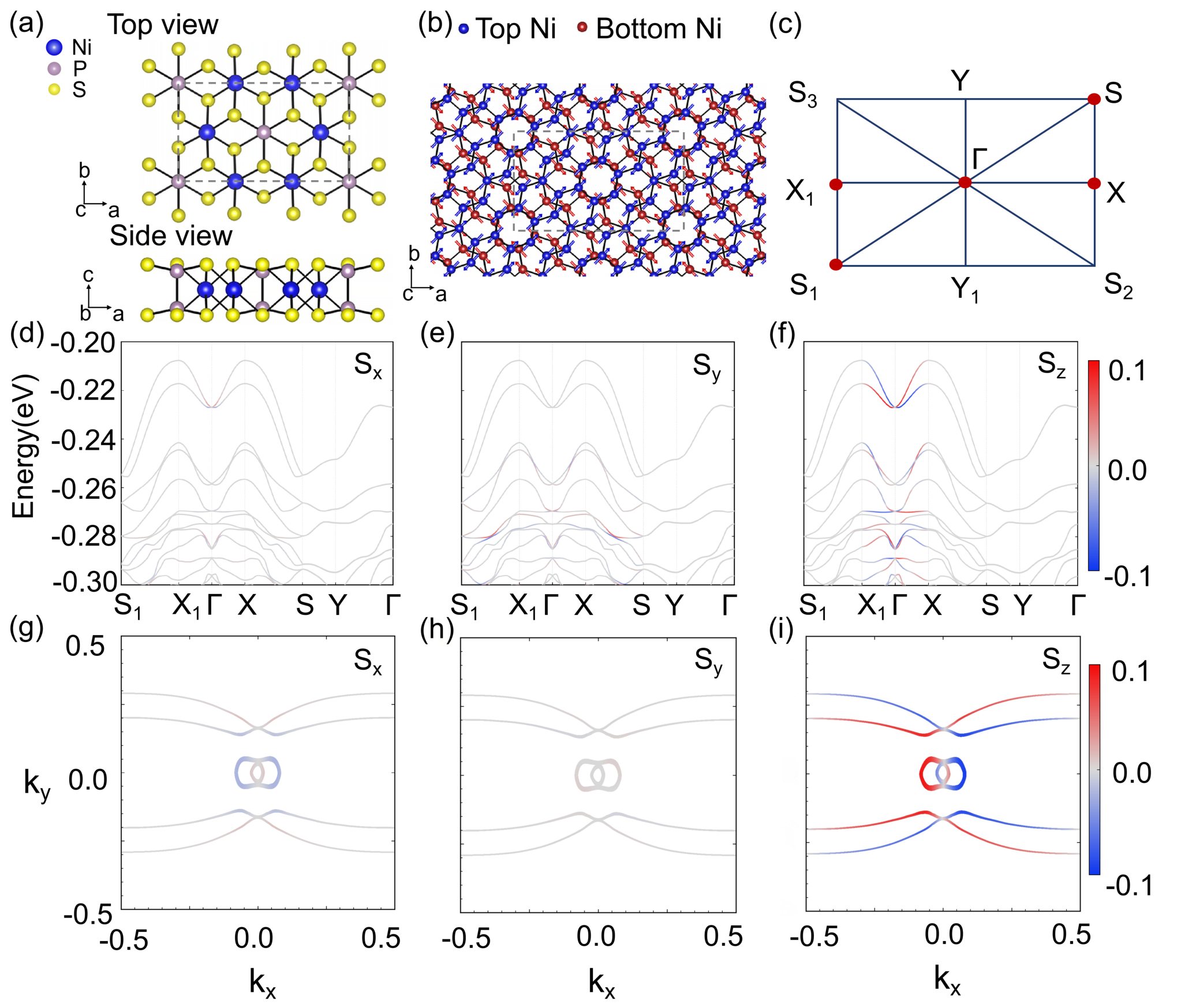}
    \caption{Atomic structure and $p$-wave magnetism in $21^\circ$-twisted bilayer $\mathrm{NiPS}_3$. (a) Top and side views of the monolayer $\mathrm{NiPS}_3$ structure.
    (b) Top view of the $21^\circ$ moir\'e superlattice; blue and red denote the top and bottom layers.
    (c) Moir\'e Brillouin zone and high-symmetry points.
    (d)--(f) Bands along high-symmetry paths, colored by the $S_x$, $S_y$, and $S_z$ projections, respectively; the split bands are polarized predominantly out of plane.
    (g)--(i) Isoenergy contours colored by the corresponding spin projections. Panel (i) shows the odd-parity $p$-wave texture $S_z(\mathbf{k})=-S_z(-\mathbf{k})$, with red (spin-up) and blue (spin-down) lobes antisymmetric about $\Gamma$.}
    \label{fig:2}
\end{figure}

\emph{$p$-wave magnetism in twisted $\mathrm{NiPS}_3$.---}The van der Waals semiconductor $\mathrm{NiPS}_3$, with its predominantly in-plane zigzag antiferromagnetic order, provides a material platform for investigating the proposed mechanism~\cite{wildes2015,leon2025,chen2024ni}. Ni atoms form a honeycomb lattice [Fig.~\ref{fig:2}(a)], with moments aligned ferromagnetically along each zigzag chain and antiferromagnetically between neighboring chains. Our monolayer total-energy calculations identify zigzag order as the lowest-energy state among the ferromagnetic, N\'eel, stripy, and zigzag configurations examined (Supplemental Material II). For both the monolayer and the untwisted bilayer, the band dispersions calculated with and without spin--orbit coupling (SOC) are nearly identical (Supplemental Material II). We perform the twisted-bilayer calculations without SOC to isolate the nonrelativistic contribution of the magnetic order. A $21^\circ$ twist produces the moir\'{e} supercell shown in Fig.~\ref{fig:2}(b). In the self-consistent noncollinear calculations, the Ni moment directions are allowed to relax without constraints. In the converged state considered here, the layer N\'eel vectors are nonparallel and lie in a common plane.

The nonrelativistic symmetry operations and their actions are summarized in Table~\ref{tab:em-symmetry} of the End Matter; Table~S4 of the Supplemental Material gives their full matrices. They include the antiunitary half-translation $\mathcal{A}=\mathcal{T}\{E|\boldsymbol{\tau}\}$, with $\boldsymbol{\tau}=(\tfrac12,\tfrac12,0)$ in fractional coordinates, and the layer-exchanging operation $\mathcal{B}=[C_{2y}^{s}\Vert C_{2y}]$. In the ideal symmetry limit, $\mathcal{A}$ enforces magnetic compensation and pairs equal-energy states at $\mathbf{k}$ and $-\mathbf{k}$ with opposite spin expectation values. At generic $\mathbf{k}$, these partners occur at different momenta, so this symmetry does not require a degeneracy at the same momentum. At $\Gamma$, by contrast, $\mathcal{A}^{2}=-1$ protects a twofold degeneracy. The four operations whose spatial part is $C_{2y}$ (operations 5--8 in Table~\ref{tab:em-symmetry}) exchange the two layers. In particular, $\mathcal{B}$ constrains the leading $p$-wave splitting to the $k_x$ direction at zero field, providing an orientational constraint distinct from the opposite-momentum pairing enforced by $\mathcal{A}$.

Along the high symmetry path $S_1$--$X_1$--$\Gamma$--$X$--$S$--$Y$--$\Gamma$ shown in Fig.~\ref{fig:2}(c), the bands split near $\Gamma$ into branches with opposite $S_z$ projections. The opposite-spin branches exchange their energy ordering between the $\Gamma$--$X$ and $\Gamma$--$X_1$ directions, corresponding to $+k_x$ and $-k_x$ [Fig.~\ref{fig:2}(f)]. The spin contrast is predominantly in the $S_z$ channel [Figs.~\ref{fig:2}(d)--\ref{fig:2}(i)]. For the assumed magnetic symmetry, the unitary operation $\mathcal{G}=\{C_{2z}^{s}\Vert E|\boldsymbol{\tau}\}$ (operation 3 in Table~\ref{tab:em-symmetry}) leaves $\mathbf{k}$ unchanged and reverses $S_x$ and $S_y$. In the ideal nonrelativistic symmetry limit, the total in-plane spin expectation values therefore vanish for nondegenerate eigenstates, while spin polarization normal to the plane of the local moments remains allowed~\cite{hellenes2023,songz2025}. This fixed polarization axis does not imply conservation of the physical spin operator $S_z$, since the coplanar noncollinear exchange field does not commute with it~\cite{hellenes2023,chakraborty2025nree,ma2026hyperspin}. We evaluate the spin-current response separately using the full spinor Wannier Hamiltonian.

The isoenergy contours near the crossing at $\Gamma$ provide a momentum-space signature of this state [Figs.~\ref{fig:2}(g)--\ref{fig:2}(i)]. The $S_z$ projection forms two lobes with opposite signs on either side of $\Gamma$. For symmetry-related states and continuously tracked branches, respectively,
\begin{equation}
S_z(\mathbf{k})=-S_z(-\mathbf{k}),\qquad \Delta_p(\mathbf{k})=-\Delta_p(-\mathbf{k}),
\end{equation}
where $2\Delta_p$ is the signed energy difference between the two branches defined above. At zero field, the nodal line lies along $k_x=0$, and the two-lobed pattern is consistent with a leading $p$-wave component near $\Gamma$. The polarization remains normal to the plane of the local moments and changes sign between symmetry-related states at opposite momenta. This differs from the in-plane spin winding of a Rashba texture~\cite{hellenes2023,brekke2024}. For the coplanar noncollinear state considered here, twisted bilayer $\mathrm{NiPS}_3$ therefore exhibits a $p$-wave spin texture whose odd parity and zero-field orientational locking follow from distinct symmetries, providing the starting point for electric-field control.

\begin{figure}[tbp]
    \centering
    \includegraphics[width=\columnwidth]{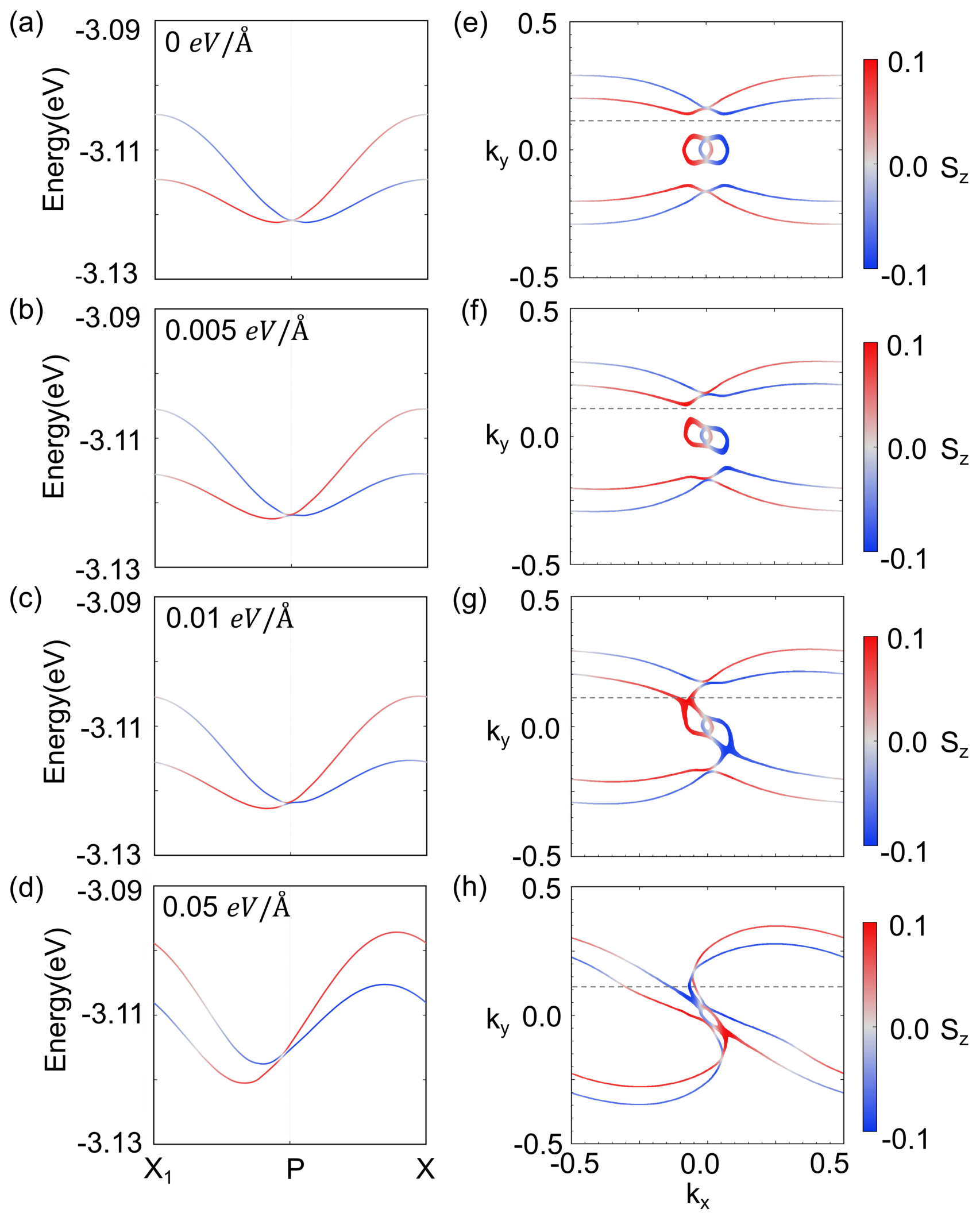}
    \caption{Electric-field-driven reorientation of the $p$-wave spin texture in $21^\circ$-twisted bilayer $\mathrm{NiPS}_3$. (a)--(d) Spin-resolved bands along $k_y=0.11$, parallel to $X_1$--$\Gamma$--$X$, for $E_\perp=0$, $0.005$, $0.01$, and $0.05$~eV/\AA{}, respectively.
    (e)--(h) Corresponding spin-resolved isoenergy contours near the crossing energy at $\Gamma$. The horizontal gray dashed lines indicate the $k_y=0.11$ cut used in (a)--(d); the central tick $P$ marks $(k_x,k_y)=(0,0.11)$.}
    \label{fig:3}
\end{figure}

\emph{Gate-tunable $p$-wave magnetism.---}A perpendicular electric field renders the two layers inequivalent. Field values quoted in eV/\AA{} denote the electron potential-energy slope $eE_\perp$, with $e>0$. We represent this effect by shifting the layer potentials by $+U_E$ and $-U_E$ relative to their mean, giving an interlayer potential-energy difference of $2U_E$. Across the range of fields examined, the converged Ni moment directions remain essentially unchanged. For a magnetic configuration retaining the in-plane antiferromagnetic translation, a uniform perpendicular field preserves operations 1--4 in Table~\ref{tab:em-symmetry}. Their spatial parts consist of the identity rotation and translations within the plane. A fixed nonzero field breaks operations 5--8, including $\mathcal{B}$, because their spatial part $C_{2y}$ reverses $z$. On the $k_y=0.11$ cut, the identity and $\mathcal{G}=\{C_{2z}^{s}\Vert E|\boldsymbol{\tau}\}$ leave $\mathbf{k}$ unchanged, whereas $\mathcal{A}=\mathcal{T}\{E|\boldsymbol{\tau}\}$ and $\mathcal{C}=[C_{2z}^{s}\Vert E]\mathcal{T}$ map this cut onto $k_y=-0.11$. The remaining symmetries therefore allow an asymmetry between $+k_x$ and $-k_x$ along this cut. We denote the point $(k_x,k_y)=(0,0.11)$ by $P$. The calculated bands become asymmetric about $P$, with their crossing shifting away from $P$ along $k_x$ [Figs.~\ref{fig:3}(b)--\ref{fig:3}(d)]. The dashed lines in the contour panels mark this cut, which avoids the symmetry-protected crossing at $\Gamma$. 


The calculated isoenergy contours evolve from a two-lobed pattern into a tilted, distorted S-shaped profile [Figs.~\ref{fig:3}(e)--\ref{fig:3}(h)]. At the contour energy selected for $0.05$~eV/\AA{}, the two pockets near $\Gamma$ merge with the outer contours. The field thus changes the orientation, shape, and connectivity of the contours. Throughout this evolution, $\mathcal{A}$ continues to enforce opposite-spin pairing at opposite momenta, $S_z(\mathbf{k})=-S_z(-\mathbf{k})$, and $\varepsilon_s(\mathbf{k},E_\perp)=\varepsilon_{-s}(-\mathbf{k},E_\perp)$ for the continuously tracked branches $s=\pm$. The odd-parity relation therefore survives the reconnection. Reciprocity of the full two-dimensional spectrum relates the $+k_y$ and $-k_y$ cuts but allows left--right asymmetry within each cut. 


Near $\Gamma$, we write the leading splitting as $v_x(E_\perp)q_x+v_y(E_\perp)q_y$. A finite transverse coefficient $v_y$ tilts the splitting vector away from the $q_x$ axis by an angle $\varphi=\operatorname{atan2}(v_y,v_x)$, tracked continuously from zero field. To linear order in momentum, the nodal line is perpendicular to this vector. Its direction measured from $q_x$ is $\varphi+\pi/2$ modulo $\pi$, so its tilt relative to the zero-field $q_y$ axis is $\varphi$. The independent curvature term $\sim q_xq_y$ also changes the contour geometry, so the full contour evolution cannot be assigned a single rotation angle. The quadratic expansion describes the local dispersion near $\Gamma$, whereas the reconnection depends on the dispersion beyond this region. For opposite field polarities, Eq.~\eqref{eq:mirror} predicts mirror-related contours of the same branch under $k_y\to-k_y$, provided the contours are evaluated at the same energy relative to a common reference.  A quantitative comparison of contour shapes and angles across fields requires a common relative-energy criterion or matched carrier densities. Within the magnetic phase considered here, the perpendicular field therefore permits reorientation and deformation while preserving the odd-parity spin pairing. Further details are provided in Sec.~III of the Supplemental Material.

\begin{figure}
    \centering
    \includegraphics[width=0.95\columnwidth]{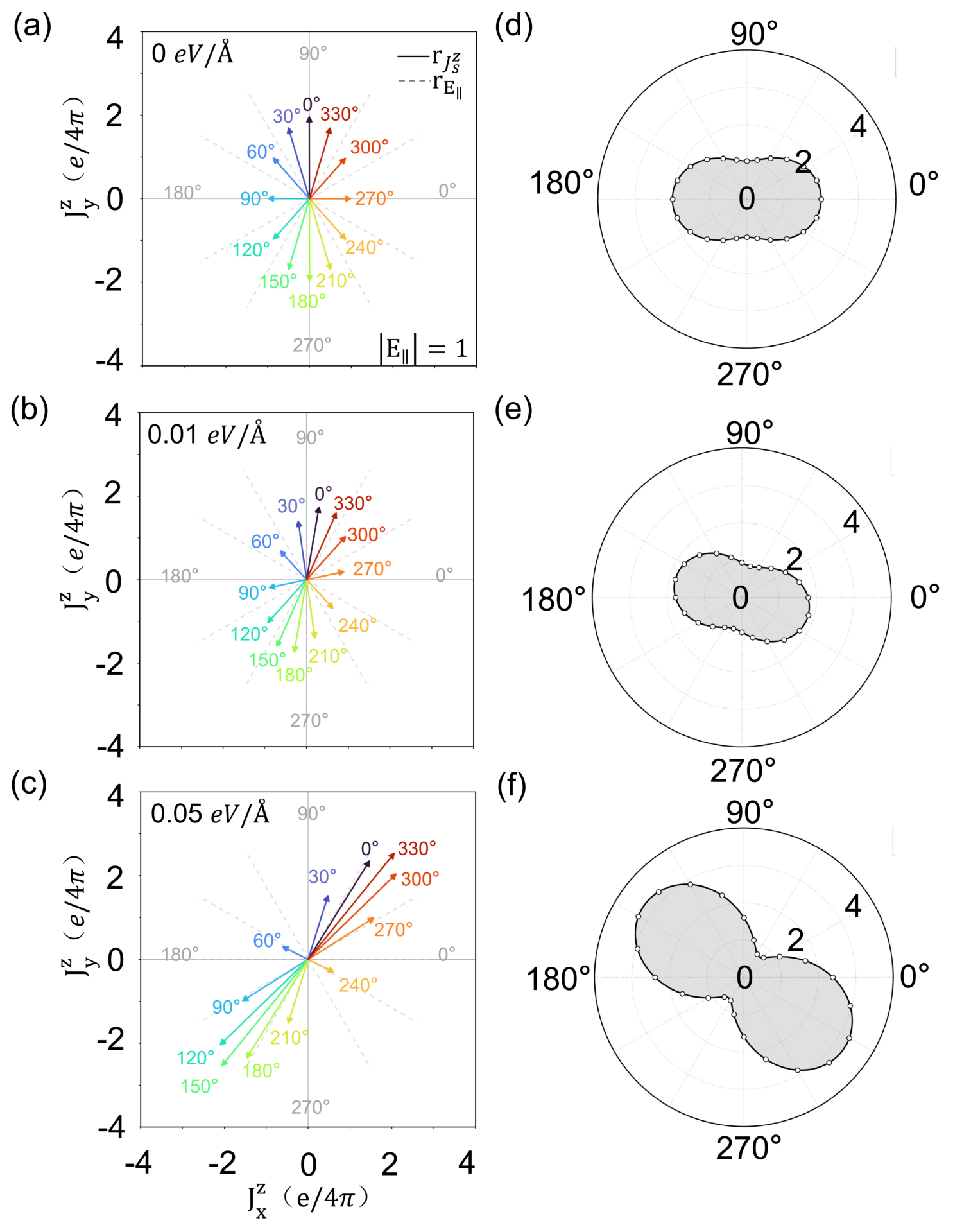}
    \caption{Angle-resolved spin-current response in $21^\circ$-twisted bilayer $\mathrm{NiPS}_3$ under perpendicular fields. (a)--(c) Spin-current vectors induced at $\varepsilon=-3.12265$~eV by $\mathbf E_\parallel=E(\cos\phi,\sin\phi)$ for $E_\perp=0$, $0.01$, and $0.05$~eV/\AA{}. Colored labels mark the driving angle $\phi$, measured from $+x$ in $30^\circ$ steps; arrows indicate current directions, and gray dashed rays indicate driving directions. The unit-drive convention $|\mathbf E_\parallel|=1$ yields normalized vectors $\mathbf J_s^z/E$.
    (d)--(f) Transverse responses from Eq.~\eqref{eq:sigmaSHeff}: the polar angle is the driving direction, the radius is $|\sigma_{\mathrm{SH,eff}}^z|$, and the signed response is positive at all displayed angles.}
    \label{fig:4}
\end{figure}

\emph{Anisotropic spin-current responses.---}We evaluate the interband $z$-polarized spin-current response using the full spinor Wannier model at the selected energy $\varepsilon=-3.12265$~eV. For an in-plane driving field $\mathbf E_\parallel=E(\cos\phi,\sin\phi)$, where $\phi$ is measured from $+x$, linear response gives
\begin{equation}
\begin{pmatrix}J_x^z\\J_y^z\end{pmatrix}
=
\begin{pmatrix}
\sigma_{xx}^z&\sigma_{xy}^z\\
\sigma_{yx}^z&\sigma_{yy}^z
\end{pmatrix}
\begin{pmatrix}E\cos\phi\\E\sin\phi\end{pmatrix},
\label{eq:Jsmatrix}
\end{equation}
where $\sigma_{ij}^z$ denotes the spin-conductivity tensor and $J_i^z$ the current flowing along $i$ and carrying the $z$ component of spin. Explicitly,
\begin{equation}
\begin{aligned}
J_x^z(\phi)&=E(\sigma_{xx}^z\cos\phi+\sigma_{xy}^z\sin\phi),\\
J_y^z(\phi)&=E(\sigma_{yx}^z\cos\phi+\sigma_{yy}^z\sin\phi).
\end{aligned}
\label{eq:Jphi}
\end{equation}
These expressions determine the vectors shown in Figs.~\ref{fig:4}(a)--\ref{fig:4}(c). The Wannier bands and spin texture are compared with first-principles results in Supplemental Material, Figs.~S7 and S8; the four tensor components are shown in Fig.~S9.

At zero perpendicular field, ideal $\mathcal B$ symmetry requires $\sigma_{xx}^z=\sigma_{yy}^z=0$. The current is transverse for drives along $x$ or $y$, while oblique drives can retain a longitudinal component when $\sigma_{xy}^z+\sigma_{yx}^z\ne0$. A finite $E_\perp$ breaks $\mathcal B$ and allows nonzero diagonal components. Over the sampled positive-field range, the diagonal entries become appreciable relative to the off-diagonal response. This changes the current vectors in Fig.~\ref{fig:4}.

To characterize the transverse response, we project the current onto $\hat{\mathbf e}_\perp=(-\sin\phi,\cos\phi)$ and define $\sigma_{\mathrm{SH,eff}}^z(\phi)
=\hat{\mathbf{e}}_\perp\cdot\mathbf{J}_s^z/E$, whose angular dependence is
\begin{equation}
\sigma_{\mathrm{SH,eff}}^z
=\sigma_{yx}^z\cos^2\phi-\sigma_{xy}^z\sin^2\phi
+\frac{\sigma_{yy}^z-\sigma_{xx}^z}{2}\sin2\phi.
\label{eq:sigmaSHeff}
\end{equation}
At zero field, $\sigma_{yx}^z>|\sigma_{xy}^z|$ and $\sigma_{xy}^z<0$ yield a positive, twofold response with maxima close to $0^\circ$--$180^\circ$ [Fig.~\ref{fig:4}(d)]. At finite field, the additional term proportional to $\sigma_{yy}^z-\sigma_{xx}^z$ tilts this angular profile [Figs.~\ref{fig:4}(e)--\ref{fig:4}(f)]. We define $\phi_0$ as the driving-field direction that maximizes the signed transverse response. At $0.05$~eV/\AA{}, the tensor gives $\phi_0=-36.3^\circ$. For a fixed drive along $+x$, the current instead flows at $\theta_J=58.1^\circ$, a change of $-32.0^\circ$ relative to its numerical zero-field direction. 

Because $\mathcal B$ reverses the perpendicular field, exchanges the layers, and reverses $S_z$, the diagonal spin-conductivity components are odd in $E_\perp$, whereas the off-diagonal components are even. The angular profiles at opposite fields are therefore related by $\phi\to-\phi$, yielding opposite tilts for a response axis tracked continuously from zero field. Layer bias thus controls the angular anisotropy and preferred axis of the spin-current response. This response-axis tilt and the texture reorientation share the loss of $\mathcal B$, but their angles need not coincide.

\emph{Conclusion and discussion.---}In summary, we have investigated electric-field control of an odd-parity spin texture in a $21^\circ$-twisted $\mathrm{NiPS}_3$ bilayer with coplanar, nonparallel layer N\'eel vectors. First-principles calculations reveal electric-field-induced reorientation and deformation of the spin-resolved isoenergy contours, while Wannier-based calculations predict tunable spin-current responses. A minimal four-band model, presented in Appendix B, illustrates how interlayer quantum interference can generate emergent $p$-wave magnetism. The nonparallel N\'eel vectors introduce spin-dependent phases into the interlayer tunneling amplitudes. Interference between distinct tunneling channels with momentum-dependent relative phases generates odd-parity spin splitting governed by the vector chirality of the two N\'eel orders, $(\mathbf{n}_1\times\mathbf{n}_2)_z$.

Layer bias therefore provides a symmetry-selective route to controlling odd-parity spin textures and anisotropic spin-current responses. In particular, the electric field breaks the layer-exchanging symmetry while preserving the antiunitary symmetry responsible for the odd-parity spin splitting, allowing the spin texture to be reoriented without changing the underlying magnetic configuration. The predicted gate-dependent angular response motivates experiments in devices that independently control carrier density and displacement field, within the range where the required magnetic state remains stable.

\emph{Acknowledgments.---}This work was supported by the National Key R\&D Program of China (Grant Nos.~2022YFB402500 and 2022YFA1403501), the National Natural Science Foundation of China (Grant Nos.~62474081, 12474174, and T2521005), and the Scientific and Technological Innovation Project of Changbaishan Laboratory, Jilin Province (Grant No.~CBS2026008). J.X.H. and L.X. acknowledge support from the Hangzhou Tsientang Education Foundation. Additional support was provided by the Max Planck Partner Group program. Calculations were performed at the high-performance computing center of Jilin University.

%

\onecolumngrid
\section*{Appendix}

\textit{Appendix A: Symmetry operations and their physical roles.---}
Table~\ref{tab:em-symmetry} lists the eight representatives of the assumed nonrelativistic spin space group, modulo magnetic lattice translations. Define $\mathcal C=[C_{2z}^{s}\mathcal T\Vert E]$ and $\mathcal G=\{C_{2z}^{s}\Vert E|\boldsymbol\tau\}$, in addition to $\mathcal A$ and $\mathcal B$ in the main text. Products in the table act from right to left; translations are given modulo a magnetic lattice vector. For a spatial rotation $R$, unitary operations send $\mathbf k$ to $R\mathbf k$, whereas antiunitary operations send it to $-R\mathbf k$. Thus a spin-only antiunitary operation need not leave momentum unchanged.

\begin{center}
\begin{minipage}{\textwidth}
\makeatletter\def\@captype{table}\makeatother
\caption{Symmetry representatives and their action on in-plane momentum, physical spin expectation values, and the perpendicular field. The numbering matches Table~S4 of the Supplemental Material. Rows 1--4 remain valid at a fixed perpendicular field if the magnetic configuration retains these symmetries; rows 5--8 relate opposite field polarities.}
\label{tab:em-symmetry}
\centering
\begin{ruledtabular}
\begin{tabular}{clccccc}
No. & Operation & Type & Translation & $(k_x,k_y)\mapsto$ & $(S_x,S_y,S_z)\mapsto$ & $E_\perp\mapsto$\\
\hline
1 & $1$ & unitary & $0$ & $(k_x,k_y)$ & $(S_x,S_y,S_z)$ & $E_\perp$\\
2 & $\mathcal C$ & antiunitary & $0$ & $(-k_x,-k_y)$ & $(S_x,S_y,-S_z)$ & $E_\perp$\\
3 & $\mathcal G$ & unitary & $\boldsymbol\tau$ & $(k_x,k_y)$ & $(-S_x,-S_y,S_z)$ & $E_\perp$\\
4 & $\mathcal A$ & antiunitary & $\boldsymbol\tau$ & $(-k_x,-k_y)$ & $(-S_x,-S_y,-S_z)$ & $E_\perp$\\
5 & $\mathcal B\mathcal C$ & antiunitary & $0$ & $(k_x,-k_y)$ & $(-S_x,S_y,S_z)$ & $-E_\perp$\\
6 & $\mathcal B$ & unitary & $0$ & $(-k_x,k_y)$ & $(-S_x,S_y,-S_z)$ & $-E_\perp$\\
7 & $\mathcal B\mathcal A$ & antiunitary & $\boldsymbol\tau$ & $(k_x,-k_y)$ & $(S_x,-S_y,S_z)$ & $-E_\perp$\\
8 & $\mathcal B\mathcal G$ & unitary & $\boldsymbol\tau$ & $(-k_x,k_y)$ & $(S_x,-S_y,-S_z)$ & $-E_\perp$\\
\end{tabular}
\end{ruledtabular}
\end{minipage}
\end{center}
\twocolumngrid

The three central constraints have distinct roles: $\mathcal A$ enforces opposite-spin pairing at opposite momenta, $\mathcal G$ excludes total in-plane spin expectation values in a nondegenerate state, and $\mathcal B$ pins the leading splitting to $k_x$ at zero field. For spin-$1/2$ electrons, $\mathcal A^2=-e^{-2i\mathbf k\cdot\boldsymbol\tau}$, which gives Kramers degeneracy at $\Gamma$. Field-induced reorientation is permitted when $\mathcal B$ is broken while $\mathcal A$ and $\mathcal G$ survive.

\textit{Appendix B: Minimal four-band model for the emergent $p$-wave magnet.---}To describe the emergence of $p$-wave magnetism and the field response of the Fermi contours, we introduce a minimal four-band $k\cdot p$
Hamiltonian in the basis $(1+,1-,2+,2-)$. This model captures the essential physics and provides a qualitative understanding of the DFT results. Although physical $S_z$ is
not conserved, $\sigma_z$ labels the conserved spin-translation sectors,
ordered so that the $s=\pm1$ sectors carry positive/negative physical $S_z$
near $\Gamma$. The Hamiltonian reads
\begin{equation}
 H_{\mathrm{eff}}=
 \begin{pmatrix}
 (\epsilon_{\rm L}+U_E+\lambda q_xq_y) I_2&T_{12}\\
 T_{12}^{\dagger}&(\epsilon_{\rm L}-U_E-\lambda q_xq_y) I_2
 \end{pmatrix},
 \label{eq:fourband}
\end{equation}
where $\mathbf q=\mathbf k-\mathbf k_\Gamma$,
$2U_E$ is the layer potential difference, and
$\epsilon_{\rm L}=E_0+A_xq_x^2+A_yq_y^2$ is the mean layer dispersion.
The opposite rotations of the layer dispersions generate the
mixed-curvature terms $\pm\lambda q_xq_y$.

Projection onto the low-energy bands gives tunneling
contributions from the even and odd combinations of the magnetic
sublattices. Their coherent sum is parameterized as
\begin{equation}
 T_{12}=a(\mathbf q)I_2+b(\mathbf q)e^{i\theta\sigma_z},
\end{equation}
with $a=g_a+i\delta_a q_x$, $b=g_b+i\delta_b q_x$, and real
$g_{a,b}$ and $\delta_{a,b}$. The angle $\theta$ measures the departure
of the N\'eel vectors from an antiparallel reference,
$\mathbf n_1\cdot\mathbf n_2=-\cos\theta$ and
$(\mathbf n_1\times\mathbf n_2)_z=\sin\theta$.
The projection and its assumptions are given in Sec.~V of the
Supplemental Material.

\begin{figure}[!t]
    \centering
    \includegraphics[width=\linewidth]{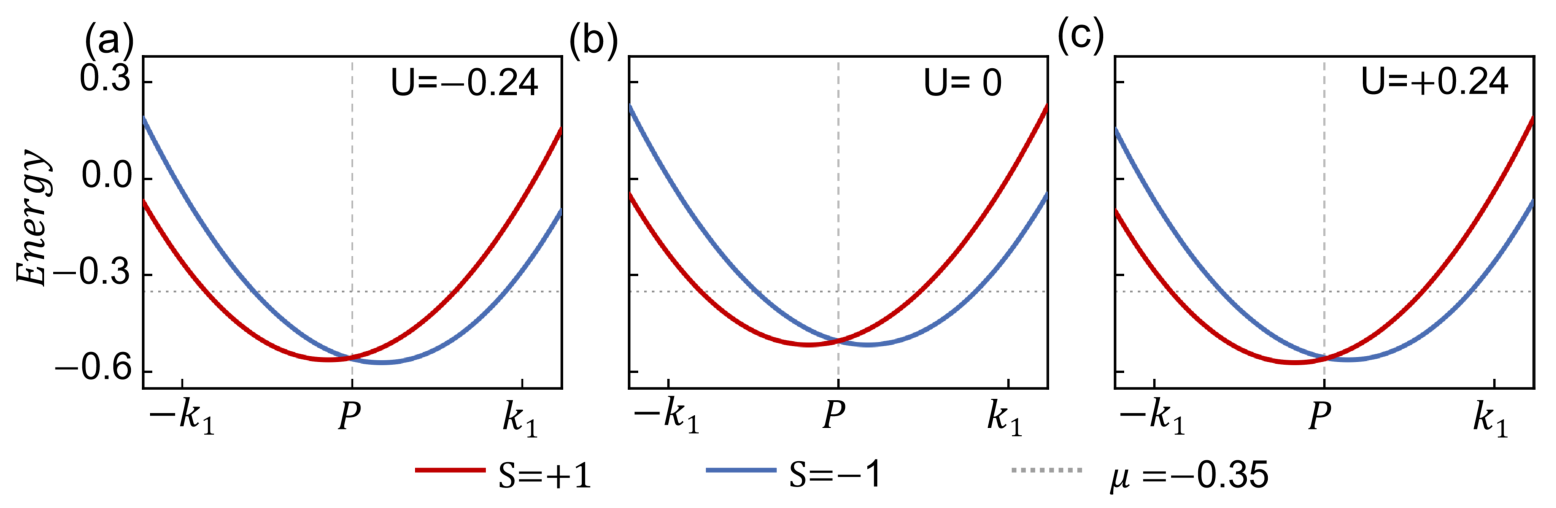}
    \caption{Lower hybridized band branches along $q_y=0.11$. Dispersions $E_{-,s}$ for (a) $U=-0.24$, (b) $U=0$, and
    (c) $U=+0.24$.
    Red and blue denote $s=+1$ and $s=-1$, respectively.
    The central tick $P$ and the vertical dashed lines mark
    $q_x=0$, corresponding to $P=(0,0.11)$.
    Horizontal dotted lines indicate $\mu=-0.35$.}
    \label{fig:em-fourband-bands}

    \vspace{1em}

    \includegraphics[width=\linewidth]{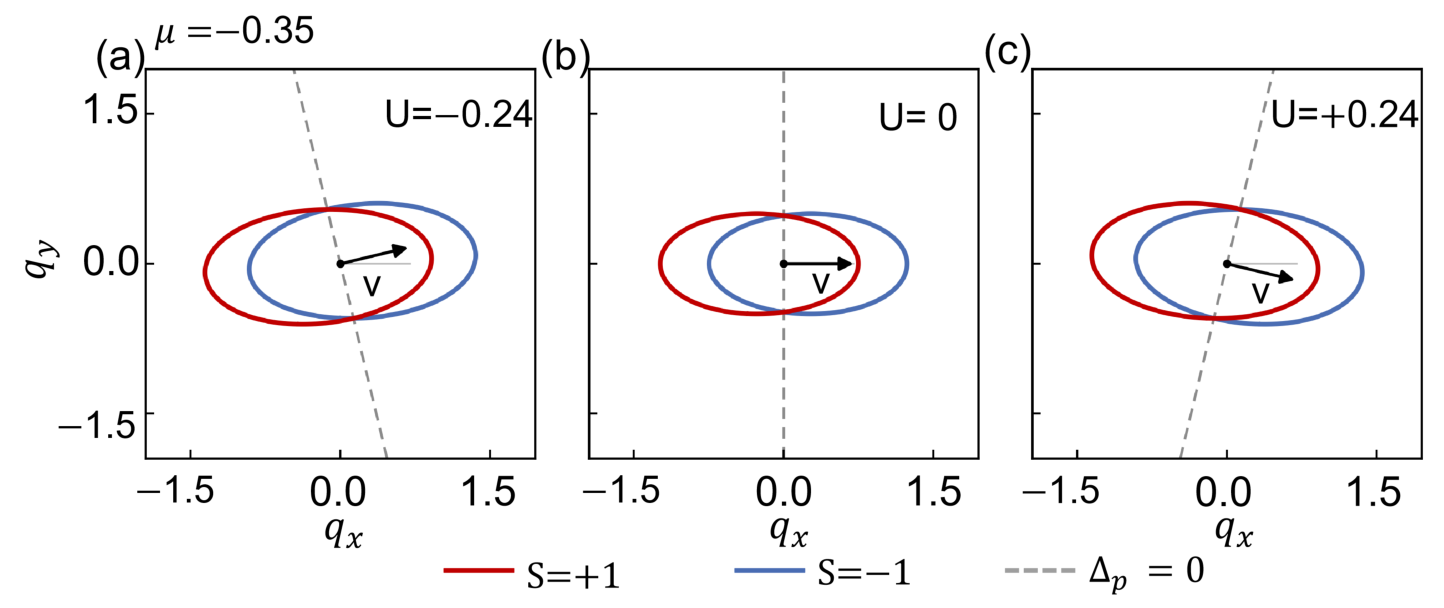}
    \caption{Bias-dependent Fermi contours. Contours of the lower hybridized pair at $\mu=-0.35$ are shown
    with the same biases and sector colors as
    Fig.~\ref{fig:em-fourband-bands}.
    Gray dashed lines mark the node $q_x+\xi Uq_y=0$.
    Black arrows indicate the leading splitting vector
    $\mathbf v$, whose angle from the positive $q_x$ axis
    is $\varphi$.}
    \label{fig:em-fourband-fermi}
\end{figure}
The antiunitary half-translation $\mathcal A$ pairs
$(\mathbf q,s,U_E)$ with $(-\mathbf q,-s,U_E)$%
~\cite{hellenes2023,jiang2024ssg}, while the layer-exchanging twofold
operation $\mathcal B$ pairs $(q_x,q_y,s,U_E)$ with
$(-q_x,q_y,-s,-U_E)$%
~\cite{sivadas2016,zhao2022zeeman,zhang2024slc}.
These operations enforce odd-parity splitting for either bias polarity
and a degeneracy line at $q_x=0$ at zero bias.
The hybridized energies are
$E_{\nu s}=\epsilon_{\rm L}+\nu\sqrt{d^2+|T_s|^2}$,
where $\nu=\pm1$, $d=U_E+\lambda q_xq_y$, and
$T_s=a+b e^{i\theta s}$. The signed splitting
$\Delta_\nu=(E_{\nu,+}-E_{\nu,-})/2$ is
\begin{equation}
 \Delta_\nu(\mathbf q,U_E)
 =-\frac{\nu\sin\theta}{D_{U}}L_xq_x
 +O(|\mathbf q|^3),
 \label{eq:split4}
\end{equation}
where
$D_{U}=[U_E^2+|g_a+g_b e^{i\theta}|^2]^{1/2}>0$
and $L_x=g_a\delta_b-g_b\delta_a$.

Equation~\eqref{eq:split4} identifies coherent interference between
two interlayer tunneling channels as the origin of the splitting:
the nonparallel N\'eel vectors supply sector-dependent phases,
while a momentum-dependent relative orbital phase produces odd
parity. Both finite vector chirality and $L_x\ne0$ are required.
A layer bias couples to the mixed curvature to tilt the Fermi
contours, while the chosen tunneling amplitudes keep the node at
$q_x=0$. 

To allow the spin-splitting node to reorient, we extend the
tunneling amplitudes to
\begin{equation}
\begin{aligned}
 a(\mathbf q,U)&=g_a+i\delta_a(q_x+\xi Uq_y),\\
 b(\mathbf q,U)&=g_b+i\delta_b(q_x+\xi Uq_y).
\end{aligned}
\label{eq:em-channels}
\end{equation}
Here $\xi$ parameterizes a symmetry-allowed bias dependence of
the tunneling beyond the scalar layer potential. 

All quantities below are dimensionless, with $U$ denoting $U_E$
and positive $U$ corresponding to a positive applied field. We assume
that the N\'eel vectors follow the lattice rotations from an
antiparallel reference, giving $\theta=21^\circ$.
We use $E_0=0$, $A_{x,y}=1/(2m_{x,y})$,
$(m_x,m_y)=(2.4,0.72)$, $(g_a,g_b)=(0.34,0.18)$,
$(\delta_a,\delta_b)=(-0.12,0.32)$, and $\xi=-1$.
The illustrative curvature parameterization is
$\lambda=(A_x-A_y)\sin21^\circ=-0.17420664$.
The chemical potential is fixed at $\mu=-0.35$ for
$U=-0.24$, $0$, and $+0.24$.
These parameters illustrate the mechanism and are not fitted
to $\mathrm{NiPS}_3$. The channel interference gives
\begin{equation}
 |T_+|^2-|T_-|^2
 =-4L_x\sin\theta\,(q_x+\xi Uq_y).
 \label{eq:em-interference}
\end{equation}
Thus, the node lies along $q_x+\xi Uq_y=0$.
For the lower pair, with
$\Delta_p=(E_{-,+}-E_{-,-})/2$, the leading splitting vector
is parallel to $(1,\xi U)$ and has $v_x>0$ for the chosen parameters.
Its angle from the positive $q_x$ axis is
$\varphi=\arctan(\xi U)$, giving $+13.50^\circ$, $0^\circ$,
and $-13.50^\circ$ at the three biases
[Fig.~\ref{fig:em-fourband-fermi}].
The pocket geometry also responds to the mixed curvature, so its
principal-axis angle need not coincide with $\varphi$.
Opposite-sector contours remain related by $\mathbf q\to-\mathbf q$,
while same-sector contours at opposite biases are related by
$q_y\to-q_y$.

Along the cut $q_y=0.11$
[Fig.~\ref{fig:em-fourband-bands}], the crossing occurs at
$q_x=-\xi U(0.11)$, giving $-0.0264$, $0$, and $+0.0264$,
respectively. This cut does not pass through $\Gamma$, where
the degeneracy persists at every bias.

\end{document}